\documentclass[aps,prd,nofootinbib,floatfix,superscriptaddress,secnumarabic]{revtex4}

\usepackage{xcolor}
\usepackage{amsmath}
\usepackage{bm}
\usepackage{graphicx}
\usepackage{epsf}
\usepackage{epsfig}
\usepackage{amsmath}
\usepackage{amsfonts,amssymb}
\usepackage{colordvi}
\usepackage{color}
\usepackage{lscape}
\usepackage{rotfloat}
\usepackage{rotating}
\usepackage{dsfont}
\usepackage{slashed,ulem}
\usepackage{booktabs}
\usepackage{rotating}
\newcommand{\be}{\begin{equation}}
\newcommand{\ee}{\end{equation}}
\newcommand{\bea}{\begin{eqnarray}}
\newcommand{\eea}{\end{eqnarray}}

\newcommand{\MSbar}{{\overline{\rm MS}}}

\newcommand{\Dslash}{{\not{\hspace{-0.1cm}D}}}

\newcommand{\la}{\lambda}

\DeclareMathOperator{\tr}{tr}
\newcommand{\trc}{\tr_c}
\def\slashed{{/}\mskip-10.0mu}

\begin{document}

\title{Perturbative renormalization of the supercurrent operator in lattice\\ ${\cal N}{=}1$ supersymmetric Yang-Mills theory}

\author{G.~Bergner}
\email[]{georg.bergner@uni-jena.de}
\affiliation{University of Jena, Institute for Theoretical Physics,\\
Max-Wien-Platz 1, D-07743 Jena, Germany}

\author{M.~Costa}
\email[]{kosta.marios@ucy.ac.cy}
\author{H.~Panagopoulos}
\email[]{panagopoulos.haris@ucy.ac.cy}
\affiliation{Department of Physics, University of Cyprus,\\
1 Panepistimiou Avenue, 2109 Aglantzia, Nicosia, Cyprus}

\author{I.~Soler}
\email[]{ivan.soler.calero@uni-jena.de}
\affiliation{University of Jena, Institute for Theoretical Physics,\\
Max-Wien-Platz 1, D-07743 Jena, Germany}

\author{G.~Spanoudes}
\email[]{g.spanoudis@cyi.ac.cy}
\affiliation{Computation-based Science
and Technology Research Center,\\
The Cyprus Institute, 20 Kavafi Str., Nicosia 2121, Cyprus}

\begin{abstract}
  In this work we perform a perturbative study of the Noether supercurrent operator in the context of Supersymmetric ${\cal N}{=}1$ Yang-Mills (SYM) theory on the lattice. The supercurrent mixes with several other operators,  some of which are not gauge invariant, having the same quantum numbers. We determine, to one loop order, the renormalization and all corresponding mixing coefficients by computing relevant Green's functions of each one of the mixing operators with external elementary fields. Our calculations are performed both in dimensional and lattice regularization. From the first regularization we obtain the $\MSbar$-renormalized Green's functions; comparison of the latter with the corresponding Green's functions in the lattice regularization leads to the extraction of the lattice renormalization factors and mixing coefficients in the $\MSbar$ scheme. The lattice calculations are performed to lowest order in the lattice spacing, using Wilson gluons and clover improved gluinos. 
  The lattice results can be used in nonperturbative studies of supersymmetric Ward identities. 

\end{abstract}

\maketitle
\section{Introduction}
 Supersymmetry (SUSY) has a variety of applications in modern quantum field theory. The most well-known are possible extensions of the standard model of particle physics and theoretical considerations like the gauge/gravity duality. In supersymmetric extensions of the standard model SUSY is expected to emerge at very high energies and it provides dark matter candidates, arising from the lightest supersymmetric particles. Supersymmetric extensions of the standard model would also resolve the hierarchy problem. In theoretical considerations, the symmetry constraints the strongly interacting state of gauge theories such that analytical predictions and conjectures are possible. These theories are typically models with extended supersymmetry such as ${\cal N} = 4$ supersymmetric Yang-Mills (SYM) theory.

In order to check and extend theoretical predictions for strongly coupled supersymmetric gauge theories, numerical investigations of lattice gauge theory would be desirable. However, it is unavoidable to break supersymmetry in any non-trivial theory on the lattice. SUSY is recovered in the continuum limit via fine tuning of the lattice theory. A signal for fine tuning is provided by supersymmetric Ward identities. However, the SUSY Ward identities usually involve a substantial mixing with different operators. It is our aim to investigate the extent to which perturbative estimates can provide some insights for this operator mixing. Our long term aim is to find strategies that might be used for theories with scalar fields like supersymmetric QCD. In these theories a large number of mixing terms appear and additional insights about mixing coefficients are essential. We will start, however, with a simpler case, where the mixing terms can be evaluated nonperturbatively. This allows to check the relevance of perturbative estimates.   

In this work, we consider the ${\cal N} = 1$ supersymmetric Yang-Mills (SYM) theory with gauge group $SU(N_c)$, which describes the strong interactions between gluons and gluinos, the superpartners of the gluons. SYM shares some of the fundamental properties of supersymmetric theories containing quarks and squarks, while at the same time it is amenable to high-accuracy nonperturbative investigations; it is thus an ideal forerunner to the future study of theories containing more superfields. 

In any SUSY theory there exists a conserved current for each generator of the SUSY algebra. The current associated with this algebra is called supercurrent. On the lattice, the supercurrent divergence is proportional to the gluino mass and  Ward identities involving the supercurrent operator are used to renormalize it nonperturbatively \cite{Ali:2018fbq}. Furthermore, other operators sharing the same quantum numbers emerge beyond the classical level. 

In this work, we use clover improved fermions and Wilson gluons to study the supercurrent operator. Its renormalization and its mixing patterns with gauge invariant and noninvariant operators will be extracted from the computation of its one-loop Green's functions with external elementary fields. Such Green's functions are not gauge independent and will be evaluated for arbitrary covariant gauge. The lattice action is chosen similar to the one used in large scale simulations \cite{Bergner:2015adz,Ali:2019agk}.

After presenting the basics of the computational setup (section \ref{comsetUP}), we calculate the renormalization of the supercurrent (section \ref{resultsALL}) both in dimensional (subsection \ref{DR}) and lattice (subsection \ref{Latt}) regularizations, using the $\MSbar$ renormalization scheme. Finally, in section~\ref{summary} we provide a short outlook.

\section{Computational setup for the renormalization of the supercurrent operator}
\label{comsetUP}

The supercurrent stems from the application of Noether's theorem to supersymmetric transformations~\cite{Martin:1997ns} of the SYM Lagrangian:
\be
{\cal L}_{\rm SYM} \to {\cal L}_{\rm SYM} + \delta_\xi {\cal L}_{\rm SYM},
\ee 
$\delta_\xi {\cal L}_{\rm SYM} = \bar \xi {\partial}_\mu Y^\mu $ and the parameter $\xi$ is a Grassmann spinor; the definition of the supercurrent is given by:
\be
\bar \xi S^\mu  = \bar \xi \left(\Sigma_i \delta_\xi \phi_i  \frac{{\partial}{\cal L}_{\rm SYM}}{{\partial}({\partial}_\mu \phi_i )} - Y^\mu \right),
\label{def}
\ee
where the index $i$ runs over all degrees of freedom ($\phi_i$) in of ${\cal L}_{\rm SYM}$.
Our studies have been performed in the Wess-Zumino (WZ) gauge. In this gauge, the SYM Lagrangian contains the gluon ($u_{\mu}$) and gluino ($\lambda$) fields, as well as an auxiliary field; the latter is eliminated, either by applying its equation of motion (classical case), or by functionally integrating over it (quantum case). Thus, the Lagrangian of SYM, in Minkowski space, becomes:
\begin{equation} 
\mathcal{L}_{\rm SYM}=-\frac{1}{4}u_{\mu \nu}^{\alpha}u_{\mu \nu}^{\alpha}+\frac{i}{2}\bar{\lambda}^{\alpha}\gamma^{\mu}\mathcal{D}_{\mu}\lambda^{\alpha}, \quad  u_{\mu \nu}= \partial_{\mu} u_\nu - \partial_{\nu} u_\mu + i g [u_\mu, u_\nu],\quad \mathcal{D}_{\mu}\lambda = \partial_{\mu}\lambda + ig [u_\mu,\lambda] 
\label{susylagr}.
\end{equation}
This definition includes the coupling constant $g$ and the field strength tensor $u_{\mu \nu} = u_{\mu \nu}^\alpha T^\alpha$ as well as the gluino field $\lambda = \lambda^\alpha T^\alpha$ are represented with generators of the $SU(N_c)$ algebra $T^\alpha$ normalized such that $\trc(T^\alpha T^\beta)=\frac12 \delta^{\alpha\beta}$.

${\cal L}_{\rm SYM}$ is invariant, up to a total derivative, under the following supersymmetric transformation:
\bea
\delta_\xi u_\mu^{\alpha} & = & -i \bar \xi \gamma^\mu \lambda^{\alpha}, \nonumber \\
\delta_\xi \lambda^{\alpha} & = & \frac{1}{4} u_{\mu \nu}^{\alpha} [\gamma^{\mu},\gamma^{\nu}] \xi. 
\label{susytransfDirac}
\eea
Given that the renormalized theory does not depend on the choice of a gauge fixing term, and given that many regularizations, in particular the lattice regularization, violate supersymmetry at intermediate steps, one may as well choose the standard covariant gauge fixing term, proportional to $(\partial_\mu u^\mu)^2$, rather than a supersymmetric variant~\cite{Miller:1983pg, Gates:1983nr}. The total SYM action thus includes the gauge-fixing term and the corresponding term involving the ghost field $c^\alpha$ which arises from the Faddeev-Popov procedure. The total action is no longer gauge-invariant, but it is Becchi-Rouet-Stora-Tyutin (BRST) invariant. The BRST transformation on all fields of the total action is as follows:
\bea
u^{\alpha}_{\mu}&\rightarrow& u^{\alpha}_{\mu}+(\partial_{\mu}c^a+g f^{\alpha \beta \gamma}c^{\beta}u^{\gamma}_{\mu})\ \eta, \nonumber \\
\lambda^{\alpha} &\rightarrow & \lambda^{\alpha} - g f^{\alpha\beta\gamma} c^{\beta} \lambda^{\gamma}\ \eta\nonumber \\
c^\alpha &\rightarrow & c^\alpha -\frac{g}{2}f^{\alpha\beta\gamma}c^\beta c^\gamma \ \eta, \nonumber \\
\bar{c}^\alpha &\rightarrow & \bar{c}^\alpha + \frac{1}{\alpha} \partial_\mu u_\mu^\alpha \ \eta,
\label{ffb}
\eea
where $f^{\alpha \beta \gamma}$ are the structure constants of $SU(N_c)$, $\eta$ is the Grassmann parameter of the BRST transformation and $\alpha$ in the last line of Eq.~(\ref{ffb}) is the gauge parameter. 

Using the transformations of Eq.~(\ref{susytransfDirac}) on ${\cal L}_{\rm SYM}$ one obtains:
\bea
Y^\mu &=& - 2i \trc (u^{\mu\,\nu} \gamma_\nu \la) - \frac{i}{4} \trc (u^{\rho\,\sigma} [\gamma_\rho,\gamma_\sigma]  \gamma^\mu \la) 
\label{Ymu}
\eea 

\medskip
From this point on, we will switch to Euclidean space, as required by the calculation of lattice Green's functions. Use of Eqs.~(\ref{def}) and~(\ref{Ymu}) leads to the supercurrent operator $S_\mu$~\cite{deWit:1975veh}, which in Euclidean space takes the form:
\be
S_\mu = -\frac{1}{2}\trc ( u_{\rho\,\sigma} [\gamma_\rho,\gamma_\sigma] \gamma_\mu \lambda ) 
\ee
The trace anomaly of this operator~\cite{Veneziano:1982ah} plays a significant role in the spectrum of SYM and it has phenomenological interest; $S_\mu$ is also involved in nonperturbative investigations of SYM on the lattice via the supersymmetric Ward identity. A proper study of $S_\mu$ must address the fact that it mixes with a number of other operators upon renormalization. These operators must have the same transformation properties under global symmetries (e.g. Lorentz, or hypercubic on the lattice, global $SU(N_c)$ transformations, ghost number, etc.) and their dimension must be lower than or equal to that of $S_\mu$. There are altogether four classes of such operators, as follows~\cite{Collins:1984xc}:
\begin{description}
\item [Class G:] Gauge-invariant operators.
\item [Class A:] BRST variation of operators.
\item [Class B:] Operators which vanish by the equations of motion.
\item [Class C:] Any other operators which respect the same global symmetries, but do not belong to the above classes; these can at most have finite mixing with $S_\mu$~\cite{Collins:1984xc}.
\end{description}

In particular, class G contains another dimension 7/2 gauge invariant operator (see Ref. \cite{Ali:2018fbq} and references therein). In the literature, it is denoted as:
\be
T_\mu = 2 \trc(u_{\mu\,\nu} \gamma_\nu \lambda) 
\ee
Exploiting the nilpotency of the BRST transformations, we determine the operators of class A. By Eq.~(\ref{ffb}), the operators must necessarily have the same index structure as $S_\mu$, i.e., one free spinor index, one Lorentz index, no free color and zero ghost number; in addition, their dimensionality must not exceed $7/2$. This requirement leaves only one candidate,~${\cal O}_{A1}$, for the class A operators:
\be
\delta_{BRST}\left( \bar c^{\alpha} \gamma_\mu \la^{\alpha}\right) =  \frac{1}{\alpha}  (\partial_\nu u_\nu^\alpha) \gamma_\mu \la^{\alpha}\eta + g f^{\alpha \beta \gamma} \bar{c}^{\alpha} c^{\beta} \gamma_\mu \la^{\gamma}\eta 
\label{BRST1op}
\ee
The operator ${\cal O}_{A1}$ is BRST invariant modulo equations of motion. In general, class A operators have vanishing matrix elements in physical external states with transverse polarization. However, they must be correctly taken into account for the renormalization of $S_\mu$. Similar comments apply to classes B and C. 

For class B operators we check the equations of motion for the gluino and gluon fields. Taking into account that operators must have zero ghost number and that the gluon equation of motion has already dimension 3, we conclude that only the gluino equation of motion may contribute; we must also multiply it by a factor of $u_\mu$ or $\slashed{u} \gamma_\mu$, and take the trace over color indices, in order to render it colorless (i.e., invariant under global $SU(N_c)$ transformations). This leads to two class B operators. 

We present all candidate gauge noninvariant operators which can mix with $S_\mu$ and belong to classes A, B, C\footnote{Operators ${\cal O}_{C5}$ and ${\cal O}_{C9}$, taken together with ${\cal O}_{A1}$, are linearly dependent; however, keeping both of them in the list affords us with additional consistency checks.}:
\bea
\label{BRSToperator}
{\cal O}_{A1} &=&  \frac{1}{\alpha} {\rm tr}_c ((\partial_{\nu} u_\nu) \gamma_\mu \la ) - ig\, {\rm{tr}}_c( [c, \bar{c}]   \gamma_\mu \la) \\[2ex]
{\cal O}_{B1} &=&  {\rm tr}_c (u_\mu \Dslash \la )\\[2ex]
{\cal O}_{B2} &=&  {\rm tr}_c (\slashed{u} \gamma_\mu \Dslash \la )\\[2ex]
{\cal O}_{C1} &=&  {\rm tr}_c (u_\mu \la) \\[2ex]
{\cal O}_{C2} &=& {\rm tr}_c (\slashed{u}  \gamma_\mu \la) \\[2ex]
{\cal O}_{C3} &=&  {\rm tr}_c (\slashed{u} \partial_{\mu} \la) \\[2ex]
{\cal O}_{C4} &=&  {\rm tr}_c ((\partial_{\mu} \slashed{u}) \, \la) \\[2ex]
{\cal O}_{C5} &=&  {\rm tr}_c ((\partial_{\nu} u_\nu) \gamma_{\mu} \la) \\[2ex]
{\cal O}_{C6} &=&  {\rm tr}_c (u_\nu \gamma_{\mu} \partial_{\nu} \la) \\[2ex]
{\cal O}_{C7} &=& i\, g \,{\rm tr}_c ([u_{\rho\,},u_{\sigma}] [\gamma_\rho,\gamma_\sigma] \gamma_\mu  \la) \\[2ex]
{\cal O}_{C8} &=& i\,g \,{\rm tr}_c ( [u_{\mu},u_{\nu}] \gamma_\nu \lambda ) \\[2ex]
{\cal O}_{C9} &=& i\,g\, {\rm tr}_c ([c, \bar{c}] \gamma_\mu \la)
\label{All_operators}
\eea
The same gauge noninvariant operators may mix with $T_\mu$ given that they share the same quantum numbers; we will also compute the renormalization factor and the mixing coefficients for $T_\mu$. The operators ${\cal O}_{C1},\,{\cal O}_{C2}$ are of lower dimension and thus they do not mix with $S_\mu$ in dimensional regularization; they may however show up on the lattice. Note that class C operators cannot contribute in the continuum for the purpose of $\MSbar$-renormalization. However, they may give finite mixing coefficients on the lattice. In producing the minimal list of mixing operators (Eqs.~(\ref{BRSToperator})-(\ref{All_operators})), we have also exploited the fact that charge conjugation ${\mathcal {C}}$ is a symmetry of the action, valid both in the continuum and lattice formulations of the theory: 
\be
{\mathcal {C}}:\left \{\begin{array}{ll}
&\hspace{-.3cm}\la^\alpha(x) T^\alpha \rightarrow - \la^\alpha(x)(T^\alpha)^*\\
&\hspace{-.3cm}c^\alpha(x) T^\alpha \rightarrow -c^\alpha(x) (T^\alpha)^*\\
&\hspace{-.3cm}\bar c^\alpha(x) T^\alpha \rightarrow - \bar c^\alpha(x) (T^\alpha)^*\\
&\hspace{-.3cm}u_\mu^\alpha(x)T^\alpha\rightarrow -u_\mu^\alpha(x)  (T^\alpha)^*\, ,\quad \mu=0,1,2,3\,.
\end{array}\right . 
\label{Chargeconjugation}
\ee

The mixing matrix is a $14 \times 14$ square block upper triangular matrix \footnote{This is due to the fact that~\cite{Collins:1984xc} class
G/A/B/C operators can mix with class (G,A,B,C)/(A,B,C)/(B,C)/(C) operators.}. The renormalized supercurrent can be written as a linear combination of these operators:
\begin{equation}
S_\mu^R  = Z_{SS} S_\mu^B +Z_{ST} T_\mu^B + Z_{SA1} {\cal O}^B_{A1} + \sum_{i=1}^{2} Z_{SBi} {\cal O}^B_{Bi}  + \sum_{i=1}^{9} Z_{SCi} {\cal O}^B_{Ci} 
\label{Zz} 
\end{equation}
Eq.~(\ref{Zz}) defines the first row of the mixing matrix. We are also interested in deriving the second row of this matrix: 
\begin{equation}
T_\mu^R  = Z_{TS} S_\mu^B + Z_{TT} T_\mu^B + Z_{TA1} {\cal O}^B_{A1} + \sum_{i=1}^{2} Z_{TBi} {\cal O}^B_{Bi}  + \sum_{i=1}^{9} Z_{TCi} {\cal O}^B_{Ci},
\label{Zzt} 
\end{equation}
where the renormalization factors are the diagonal matrix elements  $Z_{ii} = 1 + {\cal O}(g^2)\,$ and the mixing coefficients are the off-diagonal ($i \neq j$) elements $Z_{ij} = {\cal O}(g^2)$. Each $Z$ should more properly be denoted as $Z^{B,R}$ where $B$ is the regularization ($B = LR$: lattice, $DR$: dimensional regularization, etc.) and $R$ the renormalization scheme ($\MSbar$, etc.). 

\begin{figure}[ht!]
\centering
\includegraphics[scale=0.3]{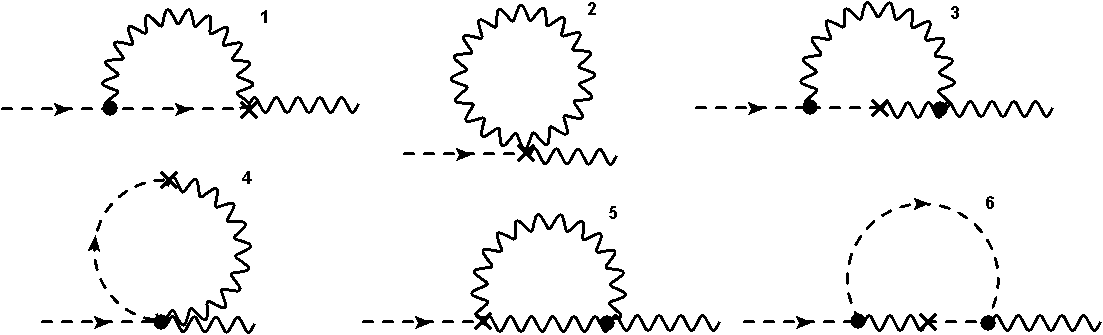}
\caption{One-loop Feynman diagrams contributing to the two-point Green's functions $\langle u_\nu S_\mu \bar \la  \rangle$\, and $\langle u_\nu T_\mu \bar \la  \rangle$.  A wavy (dashed) line represents gluons (gluinos). A cross denotes the insertion of $S_\mu$($T_\mu$). Diagrams 2, 4 do not appear in dimensional regularization; they do however show up in the lattice formulation.}
\label{fig2pt}
\end{figure}

\begin{figure}[ht!]
\centering
\includegraphics[scale=0.3]{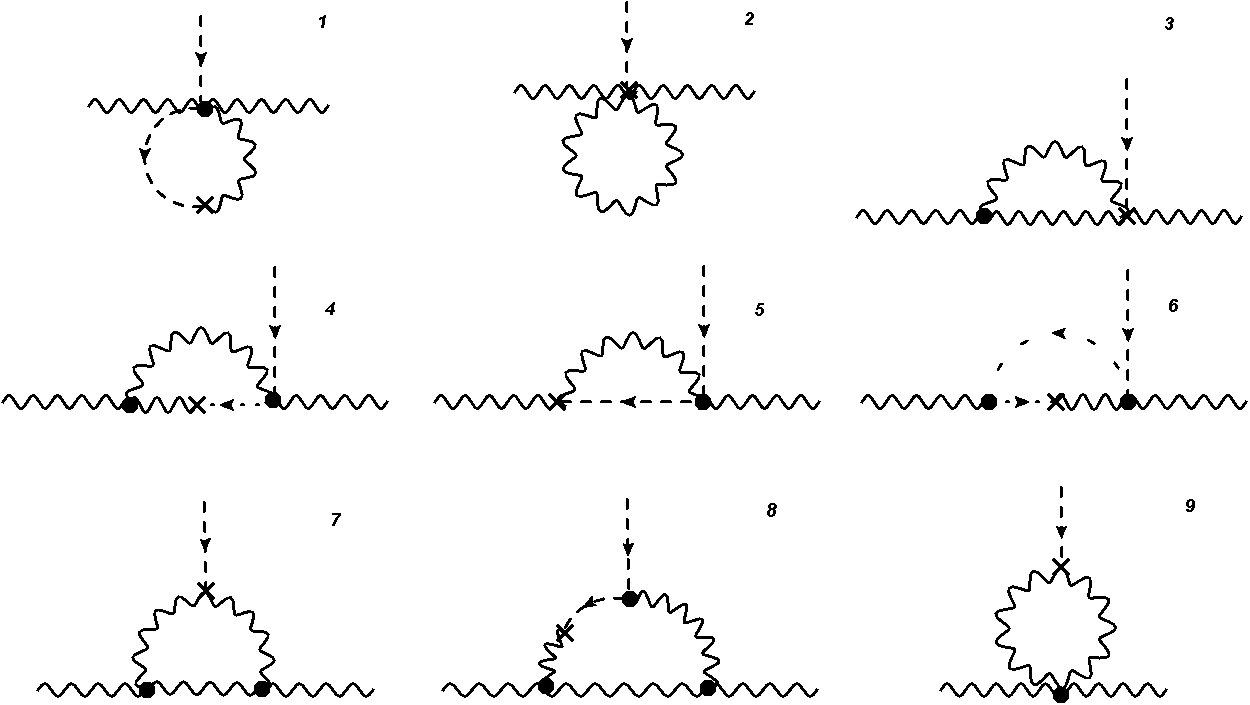}\\
\includegraphics[scale=0.3]{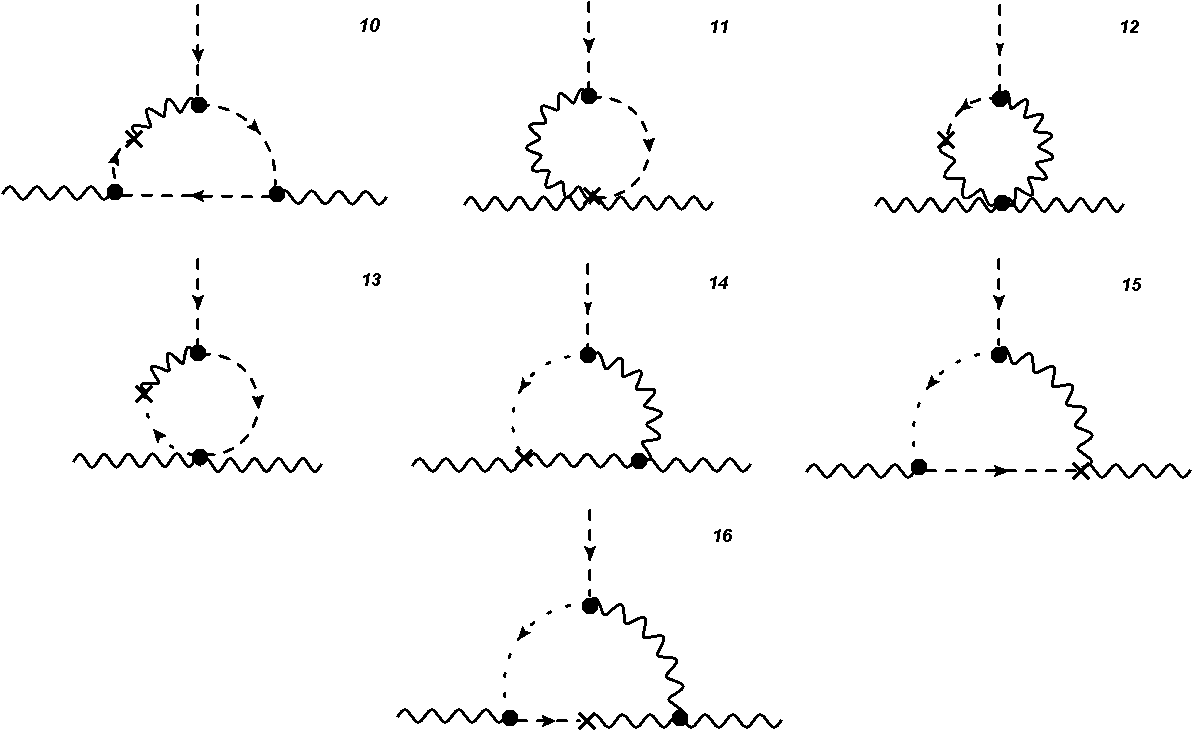}
\caption{One-loop Feynman diagrams contributing to the three-point Green's functions $\langle u_\nu u_\rho S_\mu\bar \la  \rangle$\, and $\langle u_\nu u_\rho T_\mu\bar \la  \rangle$\,.  A wavy (dashed) line represents gluons (gluinos). Diagrams 1, 2, 3, 5, 6, 11, and 13 do not appear in dimensional regularization but they contribute in the lattice regularization. A cross denotes the insertion of the operator. A mirror version (under exchange of the two external gluons) of diagrams 3, 4, 5, 6, 8, 10, 14, 15 and 16 must also be included.}
\label{fig3ptguu}
\end{figure}

\begin{figure}[ht!]
\centering
\includegraphics[scale=0.3]{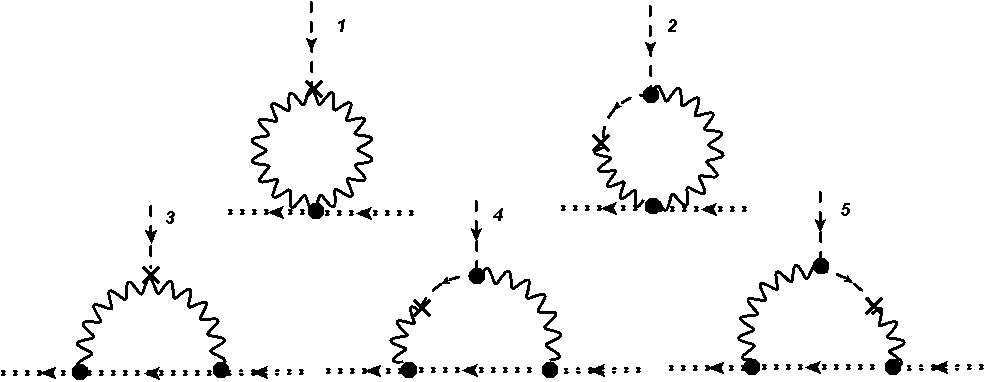}
\caption{One-loop Feynman diagrams contributing to the three-point Green's functions $\langle c \, S_\mu \, \bar c \, \bar \la \rangle$\, and $\langle c \, T_\mu \, \bar c \, \bar \la \rangle$\,.  A wavy (dashed) line represents gluons (gluinos). A cross denotes the insertion of the operator. The ``double dashed'' line is the ghost field. Diagrams 1 and 2 do not appear in dimensional regularization; they do however show up in the lattice formulation.}
\label{fig3ptgCC}
\end{figure}

In order to calculate the one-loop renormalization factors and the mixing coefficients in Eqs.~(\ref{Zz}) and~(\ref{Zzt}), we compute the two-point Green's functions of $S_\mu$ and $T_\mu$ with one external gluino and one external gluon fields (Fig.~\ref{fig2pt}), as well as three-point Green's functions with external gluino/gluon/gluon fields (Fig.~\ref{fig3ptguu}) and with external gluino/ghost/antighost fields (Fig.~\ref{fig3ptgCC}).

In gauge invariant renormalization schemes, such as the GIRS scheme~\cite{Costa:2021pfu, Costa:2021iyv}, only gauge invariant Green's functions are involved. Since the mixing of gauge-noninvariant operators leads to vanishing contributions in such Green's functions, the mixing matrix becomes effectively a $2 \times 2$ matrix which involves only $Z_{SS}, Z_{TT}, Z_{ST}, Z_{TS}$. Preliminary results for this scheme are shown in Ref.~\cite{Bergner:2021kbg}; further perturbative and nonperturbative results in the GIRS scheme and for a different discretization will be published in a follow-up paper~\cite{Bergner:2022}.

The renormalization conditions involve the renormalization factors of the gluon, gluino, ghost and coupling constant. For completeness, we present the definitions of these factors:
\bea
u_{\mu}^R &=& \sqrt{Z_u}\,u^B_{\mu},\\
\la^R &=& \sqrt{Z_\la}\,\la^B,\\
c^R &=& \sqrt{Z_c}\,c^B, \\
g^R &=& Z_g\,\mu^{-\epsilon}\,g^B, 
\eea
where $\mu$ is an arbitrary scale with dimensions of inverse length. For one-loop calculations, the distinction between $g^R$ and $\mu^{-\epsilon}\,g^B$ is inessential in many cases; we will simply use $g$ in those cases. Our results are presented as functions of the $\MSbar$ scale $\bar\mu$ which is related to $\mu$ through\footnote{$\gamma_E$  is Euler's constant: $\gamma_E = 0.57721\ldots$\, .}: $\mu = \bar \mu \sqrt{e^{\gamma_E}/ 4\pi}$.

In perturbation theory, the external fields in the Green's functions are the Fourier transformed fields and the operators are defined in position space. As is shown in Table~\ref{tb:treelevels} the tree-level Green's function with the same external fields give contributions which may depend on more than one external momentum $q_i$; this is a consequence of the absence of momentum conservation since there is no summation/integration over the position of the operators. Although this seems to complicate things it is a way to disentangle the mixing patterns. The one-loop Feynman diagrams contributing to the two-point Green's function of $S_\mu$, $\langle u_\nu S_\mu \bar \la  \rangle$, are shown in Fig.~\ref{fig2pt}. In Figs.~\ref{fig3ptguu} and~\ref{fig3ptgCC}, we present the one-loop Feynman diagrams contributing to the three-point Green's functions $\langle u_\nu  u_\rho S_\mu \bar \la  \rangle$ and $\langle c \, S_\mu \, \bar c \, \bar \la \rangle$, respectively. An analogous computation is also carried out for the corresponding Green's functions of operator $T_\mu$. Since $T_\mu$ is gauge invariant, it will be involved in the GIRS renormalization scheme; the results presented here can then be checked for consistency with the ones calculated using the GIRS scheme in Ref.~\cite{Bergner:2022}.

Previous studies on the renormalization of the supercurrent on the lattice exist in the literature~\cite{Taniguchi:1999fc, Curci:1986sm, Farchioni:2001wx, Feo:2003km, Ali:2018fbq}. In Refs.~\cite{Curci:1986sm, Farchioni:2001wx, Ali:2018fbq}, SUSY Ward identities involve the gluino mass, which receives an additive renormalization (critical mass) and the mixing with $T_\mu$ is taking into account. Ref.~\cite{Taniguchi:1999fc} investigates perturbatively the mixing behavior of the supercurrent with the on-shell condition for gluino momentum and mass using Wilson gluinos and gluons. Further, in Ref.~\cite{Hieda:2017sqq}, the gradient flow technique is used  to study the renormalization of the supercurrent in the ${\cal N}{=}1$ SYM.

\section{Results}
\label{resultsALL}

An unambiguous extraction of all mixing coefficients and renormalization constants of the operators $S_\mu$ and $T_\mu$ entails a careful selection of the appropriate Green's functions and a choice of the external momenta. In particular, we calculate two-point and three-point Green's functions of $S_\mu$ and $T_\mu$ using both dimensional regularization (continuum), where we regularize the theory in $D$-dimensions ($D = 4 - 2\epsilon$), and lattice regularization. The continuum Green's functions will be used in order to calculate the renormalized Green's functions in the $\MSbar$ scheme, which are necessary ingredients for the renormalization conditions on the lattice. 

Taking into account the potential IR divergences, we calculate the corresponding diagrams by setting to zero only one gluon or gluino external momentum. The difference between the $\MSbar$-renormalized Green's functions and the corresponding Green's functions regularized on the lattice allows us to deduce the one-loop renormalizations and mixing coefficients on the lattice. 

The novelty in our one-loop results is that we calculate the complete mixing patterns of the supercurrent operator perturbatively. More precisely, we use gauge-variant off-shell Green's functions; we obtain analytic expressions for the renormalization factors and mixing coefficients, where the number of colors, $N_c$, the coupling constant $g$, the gauge parameter, $\alpha$, the clover/Wilson parameters, $c_{\rm SW}/ r$ (on the lattice) are left unspecified.  

\subsection{Results for Green's functions and for the mixing matrix in dimensional regularization (DR)}
\label{DR}

In this subsection we present continuum results on bare Green's functions of the composite operators, $S_\mu$ and $T_\mu$, with external elementary quantum fields in the momentum space. We will use the $\MSbar$ renormalization scheme.  The divergent parts ($1/\epsilon$) of the one-loop contributions are expected to contain tensorial structures of the tree-level Green's functions of  some of the mixing operators. For this reason  we present, in Table~\ref{tb:treelevels}, the expressions for the tree-level two-point and three-point Green's functions of the operators $S_\mu$ and $T_\mu$, and of all gauge noninvariant operators which could mix with them. Notice that in Table~\ref{tb:treelevels}, the Green's functions with external gluinos, gluons and ghosts: $\langle u_\nu^{\alpha_1}(-q_1)\,{\cal O}_{i}(x)\, \bar\lambda^{\alpha_2}(q_2) \rangle_{amp}^{tree}$, $\langle u_\nu^{\alpha_1}(-q_1)\, u_\rho^{\alpha_2}(-q_2)\,{\cal O}_{i}(x)\, \bar\lambda^{\alpha_3}(q_3) \rangle_{amp}^{tree}$, $\langle c^{\alpha_3}(q_3) \,{\cal O}_{i}(x)\,\bar c^{\alpha_2}(q_2) \bar\lambda^{\alpha_1}(q_1) \rangle_{amp}^{tree}$ are shown apart from overall exponential and color factors, which are understood.
\begin{table}
\begin{center}
 \begin{tabular}{c | c | c | c} 
 \hline
 & Tree Level two-point  & Tree Level three-point & Tree Level three-point  \\
 Operators & Green's function& Green's function& Green's function \\
 & (external legs: $u_\nu\,\bar \la$) & (external legs: $u_\nu\,u_\rho \, \bar \la$)& (external legs: $c\,\bar c \, \bar \la$) \\ [0.5ex] 
 \hline\hline
$S_\mu$ & $-i (\slashed{q_1} \gamma_\nu - q_{1\nu})\gamma_\mu $  & $\,g\,[\gamma_\nu,\gamma_\rho] \gamma_\mu/2$ & $0$\\ [2ex]
 \hline
$T_\mu$ & $ i(q_{1\mu} \gamma_\nu - \slashed q_{1} \delta_{\mu \nu}  )$  & $ - \,g \, (\delta_{\mu \nu} \gamma_\rho + \delta_{\mu \rho} \gamma_\nu) $ & $0$\\ [2ex]
 \hline
 ${\cal O}_{A1}$ & $i\,q_{1\nu}\gamma_\mu/(2\alpha)$ & $0$ & $(g/2) \gamma_\mu$ \\ [2ex]
 \hline
 ${\cal O}_{B1}$ & $i\,\delta_{\mu \nu} \slashed{q_2}/2 $ & $-\,g\,(\delta_{\nu \mu} \gamma_\rho + \delta_{\rho \mu} \gamma_\nu)/2$ & $0$ \\[2ex]
 \hline
 ${\cal O}_{B2}$ & $i\,\gamma_\nu \gamma_\mu \slashed{q_2}/2 $ &  $-2\,g\,\gamma_\nu \gamma_\mu \gamma_\rho$ & $0$ \\[2ex]
 \hline
 ${\cal O}_{C1}$ & $\delta_{\mu \nu}/2$ & $0$ & $0$ \\[2ex]
 \hline
 ${\cal O}_{C2}$ & $\gamma_\nu \gamma_\mu/2$ & $0$ & $0$ \\ [2ex] 
\hline
 ${\cal O}_{C3}$ & $i\,\gamma_\nu q_{2\mu}/2$ & $0$ & $0$ \\[2ex]
 \hline
 ${\cal O}_{C4}$ & $i\,\gamma_\nu q_{1_\mu}/2$ & $0$ & $0$ \\ [2ex] 
\hline
 ${\cal O}_{C5}$ & $i\,\gamma_\mu q_{1\nu}/2$ & $0$ & $0$ \\[2ex]
 \hline
 ${\cal O}_{C6}$ & $i\,\gamma_\mu q_{2\nu}/2$ & $0$ & $0$ \\ [2ex] 
\hline
 ${\cal O}_{C7}$ & $0$ & $-g\,[\gamma_\nu,\gamma_\rho] \gamma_\mu$ & $0$ \\[2ex]
 \hline
 ${\cal O}_{C8}$ & $0$ & $-\,g\,(\delta_{\nu \mu} \gamma_\rho + \delta_{\rho \mu} \gamma_\nu)/2$ & $0$ \\ [2ex] 
\hline
 ${\cal O}_{C9}$ & $0$ & $0$ & $-(g/2) \gamma_\mu$ \\[2ex]
  \hline \hline
\end{tabular}
\caption{The two-point and three-point tree-level Green's functions of $S_\mu$ and $T_\mu$ as well as of gauge noninvariant operators which may mix with $S_\mu$.  $\langle u_\nu^{\alpha_1}(-q_1)\,{\cal O}_{i}(x)\, \bar\lambda^{\alpha_2}(q_2) \rangle_{amp}^{tree}$ and $\langle u_\nu^{\alpha_1}(-q_1)\, u_\rho^{\alpha_2}(-q_2)\,{\cal O}_{i}(x)\, \bar\lambda^{\alpha_3}(q_3) \rangle_{amp}^{tree}$ are shown apart from an overall factor of $\delta^{\alpha_{1}\alpha_{2}} e^{i\,x\cdot(q_1+q_2)}$ and $f^{\alpha_{1}\alpha_{2}\alpha_{3}} e^{i\,x\cdot(q_1+q_2+q_3)}$, respectively. Similarly, the tree-level parts of the three-point Green's functions $\langle c^{\alpha_3}(q_3) \,{\cal O}_{i}(x)\,\bar c^{\alpha_2}(q_2) \bar\lambda^{\alpha_1}(q_1) \rangle_{amp}^{tree}$ are shown apart from an overall factor of $f^{\alpha_{1}\alpha_{2}\alpha_{3}} e^{i\,x\cdot(q_1-q_2+q_3)}$.
}
\end{center}
\label{tb:treelevels}
\end{table}

We first present the continuum results of each one-loop two-point Green's function of $S_\mu$ and $T_\mu$. Use of generic values of the external momenta $q_1$ and $q_2$ lead to results which are very lengthy expressions, involving polylogarithms of the momenta; however, for the extraction of all $Z$ factors in Eqs.~(\ref{Zz}) and (\ref{Zzt}) we need only consider specific values of $q_1$ and $q_2$. In particular, a sufficient set of values consists of the following three choices: ($q_2 = 0$), ($q_1 = 0$), ($q_2 = - q_1$). For the choice $q_2=0$, we find:
 \bea
 \langle u_\nu^{\alpha_1}(-q_1)  \,{S}_{\mu}\, \bar\lambda^{\alpha_2}(q_2) \rangle _{amp}\big|^{DR}_{q_2 = 0}  &=& -i\delta^{\alpha_{1}\alpha_{2}} e^{iq_{1}x} (\slashed q_{1}\gamma_{\nu}-q_{1\nu})\gamma_\mu + i\frac{g^2 N_{c}}{16\pi^2}\frac{1}{2} \delta^{\alpha_1\,\alpha_2}e^{iq_1x}\nonumber \\ &\times& 
\bigg[
(\gamma_{\nu} \gamma_{\mu} \slashed q_1 + \gamma_\mu q_{1\nu})\left(\frac{3 (1-\alpha) }{2 \epsilon} + \frac{23}{2} + \frac{\alpha^2}{2}  + \frac{3 (1-\alpha)}{2} \log\left(\frac{\bar \mu^2}{q_1^2}\right)\right) \nonumber \\
&&- \gamma_{\nu} q_{1\mu}\left(\frac{3 (1-\alpha) }{ \epsilon} + 15 + \alpha^2 + 3(1- \alpha) \log\left(\frac{\bar \mu^2}{q_1^2}\right)\right) \nonumber \\
&& - 4 \left(\frac{\slashed q_{1} q_{1 \mu} q_{1 \nu}}{q_1^2} +\slashed q_{1} \delta_{\mu\,\nu} \right)
\bigg]
\label{GgCONTq2zero}
\eea
The above Green's function is not proportional to its tree level value, but the pole parts ($1/\epsilon$) are. In the $\MSbar$ scheme renormalization factors depend only on pole parts; removing these parts from Eq.~(\ref{GgCONTq2zero}), we are left with the $\MSbar$-renormalized Green's function. The latter will be utilized in the extraction of the lattice renormalization factors later. 
The result of the two-point Green's function for the same choice of external momentum and insertion of $T_\mu$, is:
\bea
 \langle u_\nu^{\alpha_1}(-q_1)  \,{T}_{\mu}\, \bar\lambda^{\alpha_2}(q_2) \rangle _{amp}\big|^{DR}_{q_2 = 0}  &=& -i\delta^{\alpha_{1}\alpha_{2}} e^{iq_{1}x} (q_{1\mu}\gamma_\nu - \slashed q_{1}\delta_{\mu\,\nu}) + i\frac{g^2 N_{c}}{16\pi^2}\frac{1}{2} \delta^{\alpha_1\,\alpha_2}e^{iq_1x}\nonumber \\ &\times& 
\bigg[
(\gamma_{\nu} \gamma_{\mu} \slashed q_1 + \gamma_\mu q_{1\nu})\left(   \frac{3 }{\epsilon}  + 9  + 3\log\left(\frac{\bar \mu^2}{q_1^2}\right)\right) \nonumber \\
&&- \gamma_{\nu} q_{1\mu}\left(\frac{3(1 -\alpha)}{2 \epsilon} + \frac{15}{2} + \frac{\alpha^2}{2} + \frac{3(1- \alpha) }{2} \log\left(\frac{\bar \mu^2}{q_1^2}\right)\right) \nonumber \\
&& +\slashed q_{1} \delta_{\mu\,\nu} \left(-\frac{3(3 + \alpha) }{ 2 \epsilon} -\frac{17}{2} + \alpha + \frac{\alpha^2}{2} - \frac{3(3 + \alpha) }{2} \log\left(\frac{\bar \mu^2}{q_1^2}\right)\right) 
- 2\frac{\slashed q_{1} q_{1 \mu} q_{1 \nu}}{q_1^2}  
\bigg]
\label{TGgCONTq2zero}
\eea
From Eq.~(\ref{TGgCONTq2zero}) we will calculate $Z_{TT}$ and $Z_{TS}$ as well as the mixing with other operators of Eqs.~(\ref{BRSToperator})-(\ref{All_operators}) at one-loop level.

For the choice $q_1=0$, we find:
\bea
\langle u_\nu^{\alpha_1}(-q_1)  \,{S}_{\mu}\, \bar\lambda^{\alpha_2}(q_2) \rangle _{amp}\big|^{DR}_{q_1 = 0}  &=&  i\frac{g^2 N_{c}}{16\pi^2}\frac{1}{2} \delta^{\alpha_1\,\alpha_2}e^{iq_2x}\nonumber \\ &\times& \bigg[
\gamma_{\nu} \gamma_{\mu} \slashed q_2 \left(1 - \frac{1}{2 \epsilon} - \frac{1}{2} \log\left(\frac{\bar \mu^2}{q_2^2}\right)\right) \nonumber \\
&&+ \gamma_\mu q_{2\nu} - 2 \gamma_{\nu} q_{2\mu}  - \slashed q_{2} \delta_{\mu\,\nu}\left(2 + \frac{1}{\epsilon} + \log\left(\frac{\bar \mu^2}{q_2^2}\right)\right)
\bigg]
\label{GgCONTq1zero}
\eea
Notice that the Green's function of $S_\mu$ for the choice $q_1=0$ (Eq.~(\ref{GgCONTq1zero})) is gauge independent. On the contrary, the same Green's function with the operator insertion $T_\mu$ is not.

\bea
 \langle u_\nu^{\alpha_1}(-q_1)  \,{T}_{\mu}\, \bar\lambda^{\alpha_2}(q_2) \rangle _{amp}\big|^{DR}_{q_1 = 0}  &=& i\frac{g^2 N_{c}}{16\pi^2}\frac{1}{2} \delta^{\alpha_1\,\alpha_2}e^{iq_2x}\nonumber \\ &\times& \bigg[
\gamma_{\nu} \gamma_{\mu} \slashed q_2 \left(\frac{2 + \alpha }{2 \epsilon}  + 2 + \alpha  + \frac{2+\alpha }{2} \log\left(\frac{\bar \mu^2}{q_2^2}\right)\right) \nonumber \\
&& - \gamma_\nu q_{2\mu} \left(\frac{2}{ \epsilon} +4 +\alpha + 2\log\left(\frac{\bar \mu^2}{q_2^2}\right)\right) \nonumber \\
&& +\slashed q_{2} \delta_{\mu\,\nu} \left(\frac{-2+ 2\alpha }{\epsilon}  -2 + 2\alpha + (-2+2 \alpha) \log\left(\frac{\bar \mu^2}{q_2^2}\right)\right) 
\bigg]
\label{TGgCONTTq2zero}
\eea

For the choice $q_2=-q_1$, we find:

 \bea
 \langle u_\nu^{\alpha_1}(-q_1)  \,{S}_{\mu}\, \bar\lambda^{\alpha_2}(q_2) \rangle _{amp}\big|^{DR}_{q_2 = -q_1}  &=& -i\delta^{\alpha_{1}\alpha_{2}} (\slashed q_{1}\gamma_{\nu}-q_{1\nu})\gamma_\mu  + i\frac{g^2 N_{c}}{16\pi^2}\frac{1}{2} \delta^{\alpha_1\,\alpha_2} \nonumber \\ &\times& \bigg[
 \gamma_{\nu} \gamma_{\mu} \slashed q_1 \left(\frac{4-3 \alpha }{2 \epsilon} + \frac{11}{2} + \alpha + \frac{\alpha^2}{2}  + \frac{4 - 3 \alpha }{2} \log\left(\frac{\bar \mu^2}{q_1^2}\right)\right) \nonumber \\
 && + \gamma_\mu q_{1\nu} \left(\frac{3(-1 + \alpha) }{\epsilon} -5  - 4 \alpha -\alpha^2 +  3 (-1+\alpha)  \log\left(\frac{\bar \mu^2}{q_1^2}\right)\right) \nonumber \\
&& + \slashed q_{1} \delta_{\mu\,\nu} \left(\frac{1}{\epsilon}-4 - 2\alpha +\log\left(\frac{\bar \mu^2}{q_1^2}\right)\right) \nonumber \\
&&- \gamma_{\nu} q_{1\mu}\left(\frac{3(1- \alpha) }{ \epsilon} + 5 + 4 \alpha + \alpha^2 + 3(1- \alpha) \log\left(\frac{\bar \mu^2}{q_1^2}\right)\right) \nonumber \\
&& + 4 \alpha \frac{\slashed q_{1} q_{1 \mu} q_{1 \nu}}{q_1^2}
\bigg]
\label{GgCONTq2Pq1zero}
\eea

\bea
 \langle u_\nu^{\alpha_1}(-q_1)  \,{T}_{\mu}\, \bar\lambda^{\alpha_2}(q_2) \rangle _{amp}\big|^{DR}_{q_2 = -q_1}  &=& i \delta^{\alpha_{1}\alpha_{2}}  (q_{1\mu}\gamma_\nu - \slashed q_{1}\delta_{\mu\,\nu}) + i\frac{g^2 N_{c}}{16\pi^2}\frac{1}{2} \delta^{\alpha_1\,\alpha_2}\nonumber \\ &\times& \bigg[
 \gamma_{\nu} \gamma_{\mu} \slashed q_1 \left(\frac{4 - \alpha }{2 \epsilon} + 4 + \frac{4- \alpha }{2} \log\left(\frac{\bar \mu^2}{q_1^2}\right)\right) \nonumber \\
 && + \gamma_\mu q_{1\nu} \left(\frac{3}{\epsilon} + 6 +  \alpha  + 3  \log\left(\frac{\bar \mu^2}{q_1^2}\right)\right) \nonumber \\
 &&+ \gamma_{\nu} q_{1\mu}\left(\frac{1+ 3\alpha}{2\epsilon} + \frac{1}{2}  -  \alpha - \frac{\alpha^2}{2} + \frac{1+ 3\alpha }{ 2} \log\left(\frac{\bar \mu^2}{q_1^2}\right)\right) \nonumber \\
&& + \slashed q_{1} \delta_{\mu\,\nu} \left(\frac{-5-7\alpha}{2\epsilon}-4 -\frac{13}{2} -4\alpha + \frac{\alpha^2}{2} + \frac{-5-7\alpha}{2}\log\left(\frac{\bar \mu^2}{q_1^2}\right)\right) \nonumber \\
&& + 2\alpha \frac{\slashed q_{1} q_{1 \mu} q_{1 \nu}}{q_1^2}
\bigg]
\label{GgCONTTq2Pq1zero}
\eea

In order to calculate all mixing coefficients, we also need to consider three-point Green's functions. We begin with the Green's function with two external gluon fields and one gluino field. A single choice of external momenta is sufficient in this case case: ($q_2 = 0$, $q_3 = - q_1$). The result for $S_\mu$ is\footnote{ Note the presence of $Z_g$, which is required to one loop in this case.}:
\begin{eqnarray}
\langle u_\nu^{\alpha_1}(-q_1) u_\rho^{\alpha_2}(-q_2) \,{S}_{\mu}\, \bar\lambda^{\alpha_3}(q_3) \rangle _{amp}\big|^{DR}_{q_2 =0 , q_3 = -q_1} &=& 
 \, g^R \,f^{\alpha_{1}\alpha_{2}\alpha_{3}} \frac{1}{2}[\gamma_\nu,\gamma_\rho] \,\gamma_\mu \, Z_{g}^{-1}\nonumber\\
&&\hspace{-5cm}- \frac{(g^R)^3 N_c}{16\pi^2} \,f^{\alpha_{1}\alpha_{2}\alpha_{3}}\, \Bigg[
\gamma_\nu \gamma_\rho \gamma_\mu \left(\frac{1-2\alpha}{2 \epsilon} +  \frac{5}{2} +\frac{\alpha}{4} +\frac{\alpha^2}{4} +\frac{1-2\alpha}{2} \log \left(\frac{\bar \mu^2}{q_1^2}\right)\right) \nonumber\\
&&\hspace{-5cm}+ \delta_{\nu \rho} \gamma_\mu \left(-\frac{1-2\alpha}{2 \epsilon} - \frac{29}{16} +\frac{\alpha}{8} - \frac{\alpha^2}{4} - \frac{1-2\alpha}{2} \log \left(\frac{\bar \mu^2}{q_1^2}\right)\right) + \delta_{\nu \mu} \gamma_\rho \left(\frac{19}{8} +\frac{3\alpha}{4} \right) + \delta_{\mu \rho} \gamma_\nu \left(- \frac{11}{2} +\frac{\alpha}{2} \right) \nonumber\\
&&\hspace{-5cm}+ \gamma_\nu \gamma_\mu  \frac{\slashed{q_1} q_{1\rho}}{q^2} \left( \frac{21}{16} -\frac{13\alpha}{8} -\frac{\alpha^2}{2}\right)- \gamma_\nu \gamma_\rho  \frac{\slashed{q_1} q_{1\mu}}{q^2} \left( \frac{9}{8} +\frac{\alpha}{4} \right) %
               - \gamma_\rho \gamma_\mu  \frac{\slashed{q_1} q_{1\nu}}{q^2} \left( \frac{11}{16} +\frac{\alpha}{4} \right) \nonumber\\
&&\hspace{-5cm}+  \delta_{\nu \mu} \frac{\slashed{q_1} q_{1\rho}}{q_1^2} \left(\frac{19}{8} +\frac{3\alpha}{4} \right)  
               -  \delta_{\nu \rho} \frac{\slashed{q_1} q_{1\mu}}{q_1^2} \left(\frac{5}{4} +\frac{5\alpha}{2} \right)
               + \delta_{\mu \rho} \frac{\slashed{q_1} q_{1\nu}}{q_1^2} \left(\frac{5}{8} - \frac{3\alpha}{2} \right) \nonumber\\
&&\hspace{-5cm} - \gamma_\mu \frac{q_{1\nu} q_{1\rho}}{q_1^2} \left(\frac{3}{8} +\frac{9\alpha}{4}+ +\frac{\alpha^2}{4} \right) + \gamma_\nu \frac{q_{1\rho} q_{1\mu}}{q_1^2} \left(\frac{1}{2} +\frac{9\alpha}{2}+ +\frac{\alpha^2}{2} \right) - \gamma_\rho \frac{q_{1\nu} q_{1\mu}}{q_1^2} \left(\frac{1}{4} +\frac{3\alpha}{2}+ +\frac{\alpha^2}{2} \right) \nonumber\\
+ (1 + 3\alpha)\frac{\slashed{q_1} q_{1\nu} q_{1\rho} q_{1\mu}}{q_{1}^4} 
\Bigg]
\label{threeuulS}
\end{eqnarray}

Computing the above three-point Green's function for $T_\mu$ to one loop, we find:
\begin{eqnarray}
\langle u_\nu^{\alpha_1}(-q_1) u_\rho^{\alpha_2}(-q_2) \,{T}_{\mu}\, \bar\lambda^{\alpha_3}(q_3) \rangle _{amp}\big|^{DR}_{q_2 =0 , q_3 = -q_1} &=& 
- \,g^R \, (\delta_{\mu\,\nu} \gamma_\rho + \delta_{\mu\,\rho} \gamma_\nu) \, Z_{g}^{-1}\,\nonumber\\
&&\hspace{-5cm} -\frac{(g^R)^3 N_c}{16\pi^2} \,f^{\alpha_{1}\alpha_{2}\alpha_{3}}\, \Bigg[
\gamma_\nu \gamma_\rho \gamma_\mu \left(\frac{-1+2\alpha}{4 \epsilon} +  \frac{3}{2} - \frac{\alpha}{2} - \frac{\alpha^2}{4} +\frac{-1+2\alpha}{4} \log \left(\frac{\bar \mu^2}{q_1^2}\right)\right) \nonumber\\
&&\hspace{-5cm} + \delta_{\nu \rho} \gamma_\mu \left(\frac{1-2\alpha}{4 \epsilon} + \frac{7}{8} +\frac{21\alpha}{16} + \frac{3\alpha^2}{8} + \frac{1-2\alpha}{4} \log \left(\frac{\bar \mu^2}{q_1^2}\right)\right)\nonumber\\
&&\hspace{-5cm}
+ \delta_{\nu \mu} \gamma_\rho \left(\frac{2\alpha}{\epsilon} + \frac{5}{16} - \frac{\alpha}{16} - \frac{5\alpha^2}{8}  + 2 \alpha \log \left(\frac{\bar \mu^2}{q_1^2}\right)\right)  \nonumber\\
&& \hspace{-5cm}
+ \delta_{\mu \rho} \gamma_\nu \left(-\frac{2\alpha}{\epsilon} -\frac{13}{8} +\frac{7\alpha}{8} + \frac{\alpha^2}{2} - 2 \alpha \log \left(\frac{\bar \mu^2}{q_1^2}\right)\right) \nonumber\\
&&\hspace{-5cm}+ \gamma_\nu \gamma_\mu  \frac{\slashed{q_1} q_{1\rho}}{q^2} \left( -\frac{9}{4} -\frac{19\alpha}{16} -\frac{\alpha^2}{8}\right)- \gamma_\nu \gamma_\rho  \frac{\slashed{q_1} q_{1\mu}}{q^2} \left(\frac{7}{16} +\frac{\alpha}{16} - +\frac{\alpha^2}{8} \right) %
               - \gamma_\rho \gamma_\mu  \frac{\slashed{q_1} q_{1\nu}}{q^2} \left( \frac{15}{16} - \frac{3\alpha}{4} - \frac{\alpha^2}{4} \right) \nonumber\\
&&\hspace{-5cm}+  \delta_{\nu \mu} \frac{\slashed{q_1} q_{1\rho}}{q_1^2} \left(5 +\frac{3\alpha}{8} \right)  
               -  \delta_{\nu \rho} \frac{\slashed{q_1} q_{1\mu}}{q_1^2} \left(\frac{19}{8} +\frac{21\alpha}{8} + \frac{\alpha^2}{4} \right)
               + \delta_{\mu \rho} \frac{\slashed{q_1} q_{1\nu}}{q_1^2} \left(\frac{3}{2} - \frac{21\alpha}{8} - \frac{\alpha^2}{4} \right) \nonumber\\
&&\hspace{-5cm} - \gamma_\mu \frac{q_{1\nu} q_{1\rho}}{q_1^2} \left(\frac{25}{2} +\frac{17\alpha}{2} \right) + \gamma_\nu \frac{q_{1\rho} q_{1\mu}}{q_1^2} \left(\frac{9}{2} +\frac{9\alpha}{2} +\frac{\alpha^2}{4} \right) + \gamma_\rho \frac{q_{1\nu} q_{1\mu}}{q_1^2} \left(\frac{13}{8} - \frac{13\alpha}{8} \right) \nonumber\\
+ (1 + 3\alpha)\frac{\slashed{q_1} q_{1\nu} q_{1\rho} q_{1\mu}}{q_{1}^4} 
\Bigg]
\label{threeuulT}
\end{eqnarray}

Lastly, the gluino-ghost-antighost Green's functions of the operators $S_\mu$, $T_\mu$ are shown in the following equations for a specific choice of the external momenta and they will allow us to check that the mixing coefficients $Z_{SC9}$ and $Z_{TC9}$ vanish in continuum regularization and $\MSbar$ renormalization scheme as expected. Our results for the Green's function with external gluino, antighost and ghost fields are:
\bea
\langle c^{a_3}(q_3) \,S_\mu\,\bar c^{a_2}(q_2) \bar\lambda^{a_1}(q_1) \rangle _{amp}\big|^{DR}_{q_1=q_2,\, q_3 = 0}= -\frac{g^3 N_{c}}{16\pi^2}  \,f^{\alpha_{1}\alpha_{2}\alpha_{3}}\, \alpha \left(\frac{1}{4} \gamma_\mu - \frac{\slashed{q_1} q_{1\mu}}{q_1^2}\,\right) 
\label{3ptGFgCCexpS}
\\ 
\langle c^{a_3}(q_3) \,T_\mu\,\bar c^{a_2}(q_2) \bar\lambda^{a_1}(q_1) \rangle _{amp}\big|^{DR}_{q_1=q_2,\, q_3 = 0}=  \frac{g^3 N_{c}}{16\pi^2}  \,f^{\alpha_{1}\alpha_{2}\alpha_{3}}\, \alpha \left(\frac{1}{4} \gamma_\mu + \frac{\slashed{q_1} q_{1\mu}}{2q_1^2}\,\right) 
\label{3ptGFgCCexpT}
\eea
Eqs.~(\ref{3ptGFgCCexpS}) and~(\ref{3ptGFgCCexpT}) are necessarily pole free, since ${\cal O}_{C9}$ belongs to class C. Calculation of the same Green's functions on the lattice will determine whether a (finite) mixing coefficient $z^{L,\MSbar}_{SC9}$ and $z^{L,\MSbar}_{TC9}$ will be necessary in order to match Eq.~(\ref{3ptGFgCCexpS}) and Eq.~(\ref{3ptGFgCCexpT}), respectively.

The renormalization conditions involve the renormalization factors of the external fields as well as of parameters that show up in the bare Green's functions. In the $\MSbar$ scheme, the conditions amount to the requirement that renormalized Green's functions be finite functions of the renormalized parameters, and that all contributions to $Z^{DR, \MSbar}$ factors
contain only poles in $\epsilon$. Thus, applied to the gluino-gluon Green's function of the operator ${S}_{\mu}$; the condition reads to one loop:
\bea 
\langle u_\nu^R \,{S}^R_{\mu}\, \bar\lambda^R \rangle _{amp}\Big|_{1/\epsilon = 0}=0, &\,{\rm where}\,&
\langle u_\nu^R \,{S}^R_{\mu}\, \bar\lambda^R \rangle _{amp} = Z_\la^{-1/2} \,Z_u^{-1/2}
\langle u_\nu^B \,{S}^R_{\mu}\, \bar\lambda^B \rangle _{amp}
= Z_\la^{-1/2} \,Z_u^{-1/2} Z_{SS} \langle u_\nu^B \,{S}^B_{\mu}\,\bar\lambda^B \rangle _{amp}\nonumber\\ &+& Z_{ST}  \langle u_\nu^B \, {T_\mu}^B\, \bar\lambda^B \rangle _{amp}^{tree} + Z_{SA1} \langle  u_\nu^B \, {\cal O}^B_{A1} \, \bar\lambda^B \rangle _{amp}^{tree}  + \sum_{i=1}^{2} Z_{SBi} \langle  u_\nu^B \, {\cal O}^B_{Bi} \, \bar\lambda^B \rangle _{amp}^{tree} \nonumber\\ 
&+& \sum_{i=1}^{6} Z_{SCi} \langle u_\nu^B \,{\cal O}^B_{Ci} \, \bar\lambda^B \rangle _{amp}^{tree} + {\cal O}(g^4),
\label{2ptGFexprS}
\eea
Similarly for the operator ${T}_{\mu}$:
\bea 
\langle u_\nu^R \,{T}^R_{\mu}\, \bar\lambda^R \rangle _{amp}\Big|_{1/\epsilon = 0} =0 &,& \quad \langle u_\nu^R \,{T}^R_{\mu}\, \bar\lambda^R \rangle _{amp} = Z_\la^{-1/2} \,Z_u^{-1/2} Z_{TT} \langle u_\nu^B \,{S}^B_{\mu}\,\bar\lambda^B \rangle _{amp}+ Z_{TS}  \langle u_\nu^B \, {S_\mu}^B\, \bar\lambda^B \rangle _{amp}^{tree} \nonumber\\ 
&+& Z_{TA1} \langle  u_\nu^B \, {\cal O}^B_{A1} \, \bar\lambda^B \rangle _{amp}^{tree}+ \sum_{i=1}^{2} Z_{TBi} \langle  u_\nu^B \, {\cal O}^B_{Bi} \, \bar\lambda^B \rangle _{amp}^{tree} \nonumber\\ 
&+& \sum_{i=1}^{6} Z_{TCi} \langle u_\nu^B \,{\cal O}^B_{Ci} \, \bar\lambda^B \rangle _{amp}^{tree} + {\cal O}(g^4),
\label{2ptGFexprT}
\eea

In order to determine all the above mixing coefficients, we also need to impose a set of renormalization conditions on three-point Green's functions. We will study two such Green's functions for each operator. The first one involves two external gluons and one gluino.

\be
\langle u_\nu^R u_\rho^R \,{S}^R_{\mu}\, \bar\lambda^R \rangle _{amp} \Big|_{1/\epsilon = 0} = 0, \quad  {\rm similarly\,\, for\,\,} T_\mu
\label{3ptGFexpr1S}
\ee



The second one involves external gluon, antighost and ghost fields: 
\be 
\langle c^R \,{S}^R_\mu\,\bar c^R\, \bar\lambda^R \rangle _{amp}\Big|_{1/\epsilon=0} = 0,  \quad  {\rm similarly\,\, for\,\,} T_\mu 
\label{3ptGFexpr2S}
\ee
Given that, in these cases, a power of $g^B$
appears already at tree level, the one-loop
expression for $Z_g$ must be used in renormalizing $g$; this is shown explicitly in Eqs~(\ref{threeuulS}) and (\ref{threeuulT}).
Strictly speaking, in the two-point Green's functions of Eqs.~(\ref{2ptGFexprS}),  (\ref{2ptGFexprT}) as well as the three-point Green's functions of Eqs.~(\ref{3ptGFexpr1S}) and  (\ref{3ptGFexpr2S}), one must take the regulator to its limit value (i.e. $\epsilon \to 0$ in dimensional regularization or $a \to 0$ on the lattice). This limit is convergent, provided all renormalization factors and mixing coefficients, $Z$, have been appropriately chosen.

Results for the renormalization of the external fields and of the coupling constant have been already calculated in Ref.~\cite{Costa:2020keq} and for the sake of completeness are shown for $DR$ in Eqs.~(\ref{ZuDR})-(\ref{ZgDR}).
\bea
Z_u^{DR,\MSbar} &=& 1- \frac{g^2 N_c}{16\pi^2}  \frac{1 + \alpha}{2 \epsilon} 
\label{ZuDR}\\
Z_\la^{DR,\MSbar} &=& 1+ \frac{g^2 N_c}{16\pi^2} \frac{ \alpha }{\epsilon}
\label{ZlaDR}\\
Z_{c}^{DR,\MSbar} &=& 1 - \frac{g^2 N_c}{16\,\pi^2} \frac{3-\alpha}{4\epsilon}
\label{ZcDR}\\
Z_{g}^{DR,\MSbar} &=& 1 + \frac{g^2 N_c}{16\,\pi^2} \frac{3}{2\epsilon}
\label{ZgDR}
\eea
 Imposing renormalization condition, Eq.~(\ref{2ptGFexprS}), on the two-point functions is sufficient in order to obtain the renormalization of the supercurrent $Z_{SS}$. Notice that the pole parts in Eq.~(\ref{GgCONTq2zero}) are proportional to the tree-level Green's function of $S_\mu$ and thus there are no mixing with $T_\mu$, ${\cal O}_{A1}$, ${\cal O}_{C4}$, ${\cal O}_{C5}$:  $Z_{ST} = Z_{SA1} = Z_{SC4} = Z_{SC5} = 0$. Operators ${\cal O}_{C1}$ and ${\cal O}_{C2}$ are of lower dimensionality and they will not mix in the continuum regularization: $Z_{SC1} = Z_{SC2} = 0$. By imposing the renormalization condition of Eq.~(\ref{2ptGFexprS}) and demanding the left hand side to be finite, $Z_{SS}$ is determined to be:
\be
Z_{SS}^{DR,\MSbar} = 1 + O(g^4) 
\ee
From the conceptual point of view, the case of the supercurrent is quite similar to any non-anomalous conserved current since it receives no quantum corrections in the continuum.


From Eqs.~(\ref{GgCONTq2zero}) - (\ref{3ptGFgCCexpT}) and the renormalization conditions Eqs.~(\ref{2ptGFexprS}) - (\ref{3ptGFexpr2S}) we determine to one-loop:

\bea
Z_{ST}^{DR,\MSbar} &=& 0\\
Z_{SA1}^{DR,\MSbar}  &=& 0 \\
Z_{SB1}^{DR,\MSbar} &=& \frac{g^2}{16\pi^2}  \frac{1}{\epsilon}N_c\\
Z_{SB2}^{DR,\MSbar} &=&  \frac{g^2}{16\pi^2} \frac{1}{2\epsilon} N_c\\
Z_{SCi}^{DR,\MSbar} &=& 0, \quad i=1,2,\ldots, 9  \\
Z_{TT}^{DR,\MSbar} &=& 1 - \frac{g^2}{16\pi^2} \frac{3}{\epsilon}N_c\\
Z_{TS}^{DR,\MSbar} &=& \frac{g^2}{16\pi^2}  \frac{3}{2\epsilon} N_c\\
Z_{TA1}^{DR,\MSbar}  &=& 0 \\
Z_{TB1}^{DR,\MSbar} &=& \frac{g^2}{16\pi^2}  \frac{2}{\epsilon}N_c\\
Z_{TB2}^{DR,\MSbar} &=&  -\frac{g^2}{16\pi^2} \frac{1}{2\epsilon} N_c\\
Z_{TCi}^{DR,\MSbar} &=&  0, \quad i =1,2,\ldots, 9 
\eea

As was expected, in the continuum there is no mixing with class C operators. We also see that operators $S_\mu$ and $T_\mu$ do not mix with the class A operator; however they both mix with the class B operators. Note that $T_\mu$ mixes~\cite{Abbott:1977in} with $S_\mu$, but not vice versa; its mixing coefficient $Z_{TS}$ is gauge independent in the $\MSbar$ scheme as should be for any gauge invariant operator. 


The pole-free parts of the Green's functions in Eqs.~(\ref{GgCONTq2zero}) - (\ref{3ptGFgCCexpT}) are the $\MSbar$-renormalized Green’s functions which are essential ingredients in order to extract the lattice renormalization factors and mixing coefficients.

\subsection{Results for Green's functions and for the mixing matrix in the lattice regularization (LR)}
\label{Latt}

In this subsection, we present our results for the renormalization factors and the mixing coefficients in the lattice regularization and in the $\MSbar$ scheme, as defined in the previous subsection. We make use of the Wilson formulation on the lattice, with the addition of the clover (SW) term for gluino fields. In this discretization, the Euclidean action ${\cal S}^{L}_{\rm SYM}$ on the lattice becomes:
\bea
{\cal S}^{L}_{\rm SYM}&=&a^{4}\sum_{x}\bigg[\frac{N_{c}}{g^{2}}\sum_{\mu, \nu}\bigg(1-\frac{1}{N_{c}}TrU_{\mu \nu}\bigg)+ \sum_{\mu}\bigg(Tr\bigg(\bar\lambda\gamma_{\mu} D_{\mu} \lambda\bigg)-\frac{a r}{2}Tr\bigg(\bar\lambda D^{2}\lambda \bigg)\bigg)\nonumber\\&& \hspace{5cm}-\sum_{\mu, \nu}\bigg(\frac{c_{\rm SW} \ a}{4}\bar\lambda^{\alpha}\sigma_{\mu \nu}\hat{\tilde{F}}_{\mu \nu}^{\alpha \beta}\lambda^{\beta}\bigg) + m_0 Tr \bigg(\bar\lambda \lambda\bigg)\bigg]
\label{susylagrLattice}
\eea
where
\begin{equation}
U_{\mu \nu}(x) =U_{x,x+\mu} U_{x+\mu,x+\mu+\nu} U_{x+\mu+\nu,x+\nu} U_{x+\nu,x}
\end{equation}
and $\hat{\tilde{F}}_{\mu \nu}^{ab}$ is defined in the adjoint representation  as:
\bea
\hat{\tilde{F}}_{\mu \nu}^{\alpha \beta}&=&\frac{1}{8}(\tilde{Q}_{\mu \nu}^{\alpha \beta}-\tilde{Q}_{\nu \mu}^{\alpha \beta})\\
\tilde{Q}_{\mu \nu}^{\alpha \beta}&=&2{\rm{tr}}_c \bigg( 
T^\alpha\, U_{x,x+\mu}U_{x+\mu,x+\mu+\nu}U_{x+\mu+\nu,x+\nu}U_{x+\nu,x}T^\beta\,U_{x,x+\nu}U_{x+\nu,x+\mu+\nu}U_{x+\mu+\nu,x+\mu}U_{x+\mu,x} \nonumber\\
&&\phantom{{\rm{tr}}_c } + T^\alpha\, U_{x,x+\nu}U_{x+\nu,x+\nu-\mu}U_{x+\nu-\mu,x-\mu}U_{x-\mu,x} T^\beta\, 
U_{x,x-\mu}U_{x-\mu,x-\mu+\nu}U_{x-\mu+\nu,x+\nu}U_{x+\nu,x}
\nonumber\\
&&\phantom{{\rm{tr}}_c } + T^\alpha\, U_{x,x-\mu}U_{x-\mu,x-\mu-\nu}U_{x-\mu-\nu,x-\nu}U_{x-\nu,x} T^\beta\, U_{x,x-\nu}U_{x-\nu,x-\mu-\nu}U_{x-\mu-\nu,x-\mu}U_{x-\mu,x}\nonumber\\
&&\phantom{{\rm{tr}}_c} + T^\alpha\, U_{x,x-\nu}U_{x-\nu,x-\nu+\mu}U_{x-\nu+\mu,x+\mu}U_{x+\mu,x} T^\beta\, 
U_{x,x+\mu}U_{x+\mu,x+\mu-\nu}U_{x+\mu-\nu,x-\mu}U_{x-\nu,x} \bigg)
\eea
The definitions of the covariant derivatives are as follows:
\bea
{\cal{D}}_\mu\lambda(x) &\equiv& \frac{1}{2a} \Big[ U_{x,x+\mu} \lambda (x + a\hat{\mu}) U_{x+\mu,x} - U_{x,x - \mu} \lambda(x - a \hat{\mu}) U_{x - \mu,x}) \Big] \\
{\cal D}^2 \lambda(x) &\equiv& \frac{1}{a^2} \sum_\mu \Big[ U_{x,x+\mu} \lambda (x + a \hat{\mu}) U_{x+\mu,x}  - 2 \lambda(x) +  U_{x,x -\mu} \lambda (x - a \hat{\mu}) U_{x - \mu,x}\Big]
\eea
The ``Lagrangian mass'', $m_0$, is a free parameter in principle and is related to the bare gluino mass. This term breaks supersymmetry softly. All renormalization factors which we will be calculating, must be evaluated at vanishing renormalized mass, that is, when $m_0$ is set equal to the critical value which ensures a massless gluino in the continuum limit. However, since our calculations are at one loop order this critical value is irrelevant, being already of order $g^2$. Note also that as in the continnum, a gauge-fixing term, together with the compensating ghost field term, must be added to the action, in order to avoid divergences from the integration over gauge orbits; these terms, as well as the standard ``measure'' part of the lattice action are the same as in the non-supersymmetric case~\cite{Kawai:1980ja}. The lattice analog of the BRST transformations of the continuum action is shown in Refs.~\cite{Kawai:1980ja}, \cite{Curci:1986sm}.  Further details of the lattice action can be found in Ref.\cite{Costa:2017rht}. 

Eq.~(\ref{susylagrLattice}) is invariant under the local gauge transformations:
\begin{align}
U'_{x,x+\mu} &= G^{-1}(x) U_{x,x+\mu} G(x+a \hat{\mu}), &\lambda'(x) &= G^{-1}(x) \lambda(x) G(x) &
\label{SgaugeTranComponentsL}
\end{align}
where $G(x)$ is an element of the $SU(N_c)$ gauge group in the fundamental representation. These gauge transformations commute with the lattice supersymmetry transformations (cf.~\cite{Curci:1986sm}):
\bea
\delta_\xi U_{x,x+\mu} & = & g a \, \bar \xi \gamma_\mu \lambda(x) U_{x,x+\mu}, \nonumber \\
\delta_\xi \lambda(x) & = & \frac{1}{4} [\gamma_{\mu},\gamma_{\nu}] U_{\mu \nu}(x) \xi /(i g a^2) 
\label{susytransfDiracL}
\eea

The definitions of the operators on the lattice take the form:
\be
S_\mu = -\frac{1}{2}{\rm tr}_c ( \hat{F}_{\rho \sigma} [\gamma_\rho,\gamma_\sigma] \gamma_\mu \lambda ),
\quad T_\mu = 2 {\rm tr}_c(\hat{F}_{\mu \nu} \gamma_\nu \lambda)
\ee
where
\begin{equation}
\begin{split}
\hat{F}_{\mu \nu}&=\frac{1}{8ig}(Q_{\mu \nu}-Q_{\nu \mu})\\
Q_{\mu \nu}&=U_{x,x+\mu}U_{x+\mu,x+\mu+\nu}U_{x+\mu+\nu,x+\nu}U_{x+\nu,x}\\
\quad& +U_{x,x+\nu}U_{x+\nu,x+\nu-\mu}U_{x+\nu-\mu,x-\mu}U_{x-\mu,x}\\
\quad& +U_{x,x-\mu}U_{x-\mu,x-\mu-\nu}U_{x-\mu-\nu,x-\nu}U_{x-\nu,x}\\
\quad& +U_{x,x-\nu}U_{x-\nu,x-\nu+\mu}U_{x-\nu+\mu,x+\mu}U_{x+\mu,x}
\end{split}
\end{equation} 

Both $\MSbar$-renormalized and bare Green's functions have the same tensorial structures. As is expected by renormalizability, the difference between the one-loop $\MSbar$- renormalized Green's functions (Eqs.~(\ref{GgCONTq2zero})-(\ref{3ptGFgCCexpT}), with $1/\epsilon \to 0$) and the corresponding bare lattice Green's functions must be polynomial in the external momenta. The resulting expressions for the difference between the two-point $\MSbar$-renormalized and lattice bare Green's functions of $S_\mu$ are given below in Eqs.~(\ref{firstdiff}) -(\ref{lastdiff}). Note that in all lattice expressions the systematic errors, coming from an extrapolation to infinite lattice size of our numerical loop integrals, are smaller than the last digit we present.

\bea
\label{firstdiff}
\langle u_\nu^{\alpha_1}(-q_1)  \,{S}_{\mu}\, \bar\lambda^{\alpha_2}(q_2) \rangle _{amp}\big|^{\MSbar}_{q_1 = 0}  - \langle u_\nu^{\alpha_1}(-q_1)  \,{S}_{\mu}\, \bar\lambda^{\alpha_2}(q_2) \rangle _{amp}\big|^{LR}_{q_1 = 0} &=&  i\frac{g^2 N_{c}}{16\pi^2}\frac{1}{2} \delta^{\alpha_1\,\alpha_2}e^{iq_2x}\times \Bigg[\nonumber \\ 
&& \hspace{-5cm}
\gamma_{\nu} \gamma_{\mu} \slashed q_2 \bigg(0.80802   - \frac{1}{2} \log\left(a^2 \bar \mu^2\right) \bigg) - \slashed q_{2}  \delta_{\mu\,\nu} \bigg(0.38395+ \log\left(a^2 \bar \mu^2\right) \bigg)  \Bigg]
\\
\langle u_\nu^{\alpha_1}(-q_1)  \,{S}_{\mu}\, \bar\lambda^{\alpha_2}(q_2) \rangle _{amp}\big|^{\MSbar}_{q_2 = 0}  - \langle u_\nu^{\alpha_1}(-q_1)  \,{S}_{\mu}\, \bar\lambda^{\alpha_2}(q_2) \rangle _{amp}\big|^{LR}_{q_2 = 0} &=&
i\frac{g^2}{16\pi^2}\frac{1}{2} \delta^{\alpha_1\,\alpha_2} e^{iq_{1}x} \times \Bigg[\nonumber \\
&&\hspace{-5cm} \frac{39.47842}{N_c} (\gamma_{\nu} \gamma_{\mu} \slashed q_1 + \gamma_\mu q_{1\nu})- \frac{78.95683}{N_c} \gamma_{\nu} q_{1\mu} +N_c\bigg( -5.99999 \slashed q_{1}  \delta_{\mu\,\nu} +\gamma_{\nu} q_{1\mu} 5.99722\nonumber \\
&& \hspace{-5cm}+ (\gamma_{\nu} \gamma_{\mu} \slashed q_1 + \gamma_\mu q_{1\nu} - 2 \gamma_{\nu} q_{1\mu} )\big(-30.57429 + 5.17830\alpha -  4.55519 c_{\rm SW}^2 + 5.3771 c_{\rm SW} r \nonumber \\
&& + \frac{3}{2} (1- \alpha) \log\left(a^2 \bar \mu^2\right)\big) \bigg) \Bigg]\\
\langle u_\nu^{\alpha_1}(-q_1)  \,{S}_{\mu}\, \bar\lambda^{\alpha_2}(q_2) \rangle _{amp}\big|^{\MSbar}_{q_2 = -q_1}  - \langle u_\nu^{\alpha_1}(-q_1)  \,{S}_{\mu}\, \bar\lambda^{\alpha_2}(q_2) \rangle _{amp}\big|^{LR}_{q_2 = -q_1} &=&
i\frac{g^2}{16\pi^2}\frac{1}{2} \delta^{\alpha_1\,\alpha_2}  \times \Bigg[\nonumber \\
&&\hspace{-5cm}\frac{39.47842}{N_c} (\gamma_{\nu} \gamma_{\mu} \slashed q_1 + \gamma_\mu q_{1\nu}) - \frac{78.95683}{N_c} \gamma_{\nu} q_{1\mu} + N_c\bigg( 0.80802 \gamma_\mu q_{1\nu} + 4.38396\gamma_\nu q_{1\mu}\nonumber \\
&& \hspace{-5cm}
+\frac{1}{2} \gamma_{\nu} \gamma_{\mu} \slashed q_1\log\left(a^2 \bar \mu^2\right)
\nonumber \\
&& \hspace{-5cm}+ (\gamma_{\nu} \gamma_{\mu} \slashed q_1 + \gamma_\mu q_{1\nu}- 2\gamma_\nu q_{1\mu} ) \big(-31.38231 + 5.17830 \alpha - 4.55519 c_{\rm SW}^2 + 5.37708 c_{\rm SW} r \nonumber \\
&& \hspace{-5cm} + 2 \log\left(a^2 \bar \mu^2\right) -  \frac{3}{2} \alpha\log\left(a^2 \bar \mu^2\right) \big)\nonumber \\
&&\hspace{-1cm}+\slashed q_{1} \delta_{\mu\,\nu} \left(-5.61605 +\log\left(a^2 \bar \mu^2\right) \right)
\bigg)
\Bigg]
\label{lastdiff}
\eea

Similarly, the expressions for the Green's functions with an external gluon and gluino fields and operator insertion of $T_\mu$ are as follows:

\bea
\langle u_\nu^{\alpha_1}(-q_1)  \,{T}_{\mu}\, \bar\lambda^{\alpha_2}(q_2) \rangle _{amp}\big|^{\MSbar}_{q_1 = 0}  - \langle u_\nu^{\alpha_1}(-q_1)  \,{T}_{\mu}\, \bar\lambda^{\alpha_2}(q_2) \rangle _{amp}\big|^{LR}_{q_1 = 0} &=&  i\frac{g^2 N_{c}}{16\pi^2}\frac{1}{2} \delta^{\alpha_1\,\alpha_2}e^{iq_2x}\times \Bigg[\nonumber \\ 
&& \hspace{-5cm}
\gamma_{\nu} \gamma_{\mu} \slashed q_2 \bigg(0.19198 + \frac{1}{2} \log\left(a^2 \bar \mu^2\right) \bigg) + \slashed q_{2}  \delta_{\mu\,\nu} \bigg(0.23209 -2 \log\left(a^2 \bar \mu^2\right) \bigg)  \Bigg]
\\
\langle u_\nu^{\alpha_1}(-q_1)  \,{T}_{\mu}\, \bar\lambda^{\alpha_2}(q_2) \rangle _{amp}\big|^{\MSbar}_{q_2 = 0}  - \langle u_\nu^{\alpha_1}(-q_1)  \,{T}_{\mu}\, \bar\lambda^{\alpha_2}(q_2) \rangle _{amp}\big|^{LR}_{q_2 = 0} &=&
i\frac{g^2}{16\pi^2}\frac{1}{2} \delta^{\alpha_1\,\alpha_2} e^{iq_{1}x} \times \Bigg[\nonumber \\
&&\hspace{-5cm} \frac{39.47842}{N_c} (\gamma_{\nu} \gamma_{\mu} \slashed q_1 + \gamma_\mu q_{1\nu})\nonumber \\
&& \hspace{-12cm}+N_c\bigg( -41.73349 \slashed q_{1}  \delta_{\mu\,\nu} + (\gamma_{\nu} \gamma_{\mu} \slashed q_1 + \gamma_\mu q_{1\nu})\left(4.07960 - 4.57771 c_{\rm SW}^2 + 5.73984 c_{\rm SW} r +3 \log\left(a^2 \bar \mu^2\right)\right)\nonumber \\
&&\hspace{-10cm} +\gamma_{\mu} q_{1\nu} \left(4.07960 - 4.57771 c_{\rm SW}^2 - 3.73984 c_{\rm SW} r  +3 \log\left(a^2 \bar \mu^2\right)\right)\nonumber \\
&&\hspace{-10cm} +\gamma_{\nu} q_{1\mu}\left(33.57429  + 5.17830 \alpha + 4.55519 c_{\rm SW}^2 - 5.37708 c_{\rm SW} r - \frac{3}{2} (1+\alpha) \log\left(a^2 \bar \mu^2\right)\right) 
\bigg)
\Bigg]\\
\langle u_\nu^{\alpha_1}(-q_1)  \,{T}_{\mu}\, \bar\lambda^{\alpha_2}(q_2) \rangle _{amp}\big|^{\MSbar}_{q_2 = -q_1}  - \langle u_\nu^{\alpha_1}(-q_1)  \,{T}_{\mu}\, \bar\lambda^{\alpha_2}(q_2) \rangle _{amp}\big|^{LR}_{q_2 = -q_1} &=&
i\frac{g^2}{16\pi^2}\frac{1}{2} \delta^{\alpha_1\,\alpha_2} \times \Bigg[\nonumber \\
&&\hspace{-5cm}\frac{39.47842}{N_c} (\slashed q_1  \delta_{\mu\,\nu} + \gamma_\nu q_{1\mu}) \nonumber \\
&& \hspace{-12cm}+N_c\bigg( \gamma_{\nu} \gamma_{\mu} \slashed q_1 \left(3.88762 - 4.57771 c_{\rm SW}^2 -3.73984 c_{\rm SW} r + \frac{5}{2} \log\left(a^2 \bar \mu^2\right) -  \frac{3}{2} \alpha\log\left(a^2 \bar \mu^2\right) \right)\nonumber \\
&&\hspace{-10cm}+\slashed q_{1} \delta_{\mu\,\nu} \left(-41.96558 +5.17830 \alpha + 4.60023 c_{\rm SW}^2 + 12.85677 c_{\rm SW} r -\frac{1}{2}(5 + 3 \alpha)\log\left(a^2 \bar \mu^2\right) \right) \nonumber \\
&&\hspace{-10cm}+\gamma_\mu q_{1\nu}\left(4.07960 - 4.57771c_{\rm SW}^2 + 3.73984 c_{\rm SW} r + 3 \log\left(a^2 \bar \mu^2\right)\right)\nonumber \\
&&\hspace{-10cm}+\gamma_\nu q_{1\mu}\left(33.57429 - 5.17830\alpha + 9.11039 c_{\rm SW}^2 -\frac{3}{2} (1-\alpha) \log\left(a^2 \bar \mu^2\right)\right)
\bigg)
\Bigg]
\label{lastdiff2}\eea

The absence of $q$-independent terms in Eqs.(\ref{firstdiff})-(\ref{lastdiff2}) signal that the lower-dimensional operators ${\cal O}_{C1}$ and ${\cal O}_{C2}$ do not mix with either $S_\mu$ or $T_\mu$\,.

The quantities  $Z_{u}^{LR,\overline{\textrm{MS}}}, Z_{\lambda}^{LR,\overline{\textrm{MS}}}, Z_{g}^{LR,\overline{\textrm{MS}}}$ and $Z_c^{LR,\MSbar}$, appearing in the renormalization conditions on the lattice,  they have been calculated in previous works \cite{Costa:2017rht, Costa:2020keq} at one loop level. The first three renormalization factors are computed in Ref.~\cite{Costa:2020keq}; for a self-contained presentation, they are shown below:
\begin{eqnarray}
Z_{u}^{LR,\overline{\textrm{MS}}} &=&  1 + \frac{g^2\,N_c}{16\,\pi^2} \bigg[19.7392\frac{1}{N_c^2} -18.5638 + 1.3863 \alpha + 18.8508\, c_{\rm SW}^2-1.5939\, c_{\rm SW}r + \left(\frac{3}{2}+\frac{\alpha}{2}\right) \log(a^2\,\bar\mu^2)\bigg)
\label{ZuL}
\\
Z_{\lambda}^{LR,\overline{\textrm{MS}}} &=&  1 - \frac{g^2\,N_c}{16\,\pi^2} \bigg(12.8524 + 3.7920 (1-\alpha) - 5.5891\, c_{\rm SW}^2 - 4.4977\, c_{\rm SW} r + \alpha \log(a^2\,\bar\mu^2) \bigg)
\label{ZlaL}
\\
Z_{g}^{LR,\overline{\textrm{MS}}} &=& 1 + \frac{g^2\,}{16\,\pi^2}\Bigg[ -9.8696 \frac{1}{N_c}+ N_c\Big(12.8904 + 0.7969 \, c_{\rm SW}\ r - 9.4254 \, c_{\rm SW}^2 - \frac{3}{2}\log(a^2\,\bar\mu^2)\Big)\Bigg]
\label{ZgL}
\end{eqnarray}
In Ref.~\cite{Costa:2017rht} we calculate  $Z_{c}^{L,\MSbar}$, using Wilson gluinos without clover term. Since the ghost propagator does not involve gluino fields at one loop, the clover term will not affect this one-loop renormalization factor:
\be
Z_c^{LR,\MSbar} = 1 - \frac{g^2 N_c}{16\pi^2} \Bigl[3.6086 - 1.2029 \alpha -\frac{1}{4}\left( 3 - \alpha \right)
\log\left(a^2\,\bar{\mu}^2\right) \Bigr].
\ee

 Starting from the two-point Green's function with the choice $q_2=0$, we obtain the following one-loop results:
\bea
Z_{SS}^{LR,\MSbar} &=& 1 + \frac{g^2}{16\pi^2}(\frac{-9.86960}{N_c} + N_c(-2.3170 + 14.49751c_{\rm SW}^2 - 1.23662c_{\rm SW}\,r))\\
Z_{ST}^{LR,\MSbar} &=& \frac{g^2}{16\pi^2} 3N_c\\
Z_{SA1}^{LR,\MSbar}  &=& Z_{SC_4}^{LR,\MSbar} = Z_{SC5}^{LR,\MSbar}  = 0 \\
Z_{TT}^{LR,\MSbar} &=& 1 + \frac{g^2}{16\pi^2} \left(\frac{-9.86960}{N_c} + N_c \left(3.26262 + 9.91980 c_{\rm SW}^2  - 4.97646 c_{\rm SW}\,r + 3 \log\left(a^2 \bar \mu^2\right)\right)\right)\\
Z_{TS}^{LR,\MSbar} &=& \frac{g^2}{16\pi^2} N_c \left(-2.03980+ 2.28886 c_{\rm SW}^2 + 1.86992 c_{\rm SW} r -\frac{3}{2} \log\left(a^2 \bar \mu^2\right)\right)\\
Z_{TA1}^{LR,\MSbar}  &=& Z_{TC4}^{LR,\MSbar} = Z_{TC5}^{LR,\MSbar}  = 0
\eea
An important feature of the supercurrent operator is that its renormalization is finite; its mixing with $T_\mu$ on the lattice is in agreement with Ref.~\cite{Taniguchi:1999fc}, where it is mentioned that $Z_{ST}$ is related to the gamma-trace anomaly~\cite{Kaymakcalan:1983jh} corresponding to super-conformal symmetry breaking and is identical to the one loop level $\beta$-function. 

Since for the choice $q_2=0$ the tree-level two-point Green's functions of ${\cal O}_{B1}, {\cal O}_{B1}, {\cal O}_{C3}, {\cal O}_{C6}$ vanish, we evaluate the two-point Green's functions at $q_1=0$, leading to:
\bea
Z_{SB1}^{LR,\MSbar} &=& \frac{g^2}{16\pi^2} N_c \left(-0.38395 - \log\left(a^2 \bar \mu^2\right)\right)\label{ZSB1}\\
Z_{SB2}^{LR,\MSbar} &=&  \frac{g^2}{16\pi^2} N_c \left(0.80802 -\frac{1}{2} \log\left(a^2 \bar \mu^2\right)\right)\label{ZSB2}\\
Z_{SC3}^{LR,\MSbar} &=& Z_{SC6}^{LR,\MSbar}  = 0\\
Z_{TB1}^{LR,\MSbar} &=& \frac{g^2}{16\pi^2} N_c \left(0.23209 -2 \log\left(a^2 \bar \mu^2\right)\right)\label{ZTB1}\\
Z_{TB2}^{LR,\MSbar} &=&  \frac{g^2}{16\pi^2} N_c \left(0.19197 +\frac{1}{2} \log\left(a^2 \bar \mu^2\right)\right)\label{ZTB2}\\
Z_{TC3}^{LR,\MSbar} &=& Z_{TC6}^{LR,\MSbar}  = 0
\eea

All the above are consistent with the continuum by checking the pole parts and the logarithmic divergences in the lattice spacing. At this point, we check the Green's functions for the choice $q_2=-q_1$; we find agreement with the above results. 

The three-point Green's functions determine the mixing with ${\cal O}_{C7}$, ${\cal O}_{C8}$ and ${\cal O}_{C9}$. Their results are shown below:
\bea
&&\langle u_\nu^{\alpha_1}(-q_1) u_\rho^{\alpha_2}(-q_2) \,{S}_{\mu}\, \bar\lambda^{\alpha_3}(q_3) \rangle _{amp}\big|^{\MSbar}_{q_2 =0 , q_3 = -q_1} -\langle u_\nu^{\alpha_1}(-q_1) u_\rho^{\alpha_2}(-q_2) \,{S}_{\mu}\, \bar\lambda^{\alpha_3}(q_3) \rangle _{amp}\big|^{LR}_{q_2 =0 , q_3 = -q_1} =\nonumber \\
&& \hspace{1cm}
 \frac{g^3 N_c}{16\pi^2}\,f^{\alpha_{1}\alpha_{2}\alpha_{3}}\, \Bigg[
+ (\delta_{\nu \rho} \gamma_\mu-\gamma_\nu \gamma_\rho \gamma_\mu ) \big(\frac{19.73920}{N_c^2} - 12.48660 + 3.28231 \alpha - 2.27761 c_{\rm SW}^2 + 2.68854 c_{\rm SW} r \nonumber \\
&&\hspace{10cm}
+ \frac{1-2\alpha}{2} \log \left(\frac{\bar \mu^2}{q_1^2}\right)\big) 
-2 \delta_{\nu \mu} \gamma_\rho  +2 \delta_{\mu \rho} \gamma_\nu 
\Bigg]
\label{3ptGGgLattS}
\eea

\bea
&&\langle u_\nu^{\alpha_1}(-q_1) u_\rho^{\alpha_2}(-q_2) \,{T}_{\mu}\, \bar\lambda^{\alpha_3}(q_3) \rangle _{amp}\big|^{\MSbar}_{q_2 =0 , q_3 = -q_1} - \langle u_\nu^{\alpha_1}(-q_1) u_\rho^{\alpha_2}(-q_2) \,{T}_{\mu}\, \bar\lambda^{\alpha_3}(q_3) \rangle _{amp}\big|^{LR}_{q_2 =0 , q_3 = -q_1} =
\nonumber \\
&& \hspace{1cm}\frac{g^3 N_c}{16\pi^2} \,f^{\alpha_{1}\alpha_{2}\alpha_{3}}\, \Bigg[
(\gamma_\nu \gamma_\rho \gamma_\mu -\delta_{\nu \rho}  \gamma_\mu) \left(-1.84782  + 2.28886 c_{\rm SW}^2  + 1.86992 c_{\rm SW} r - \log \left(a^2 \bar \mu^2 \right)\right) \nonumber\\
&&\hspace{0.5cm}
+ (\delta_{\nu \mu} \gamma_\rho - \delta_{\mu \rho} \gamma_\nu)  \left(\frac{19.7392}{N_c^2} -17.1822 + 3.2823\alpha +  2.30011 c_{\rm sw}^2 + 6.42838 c_{\rm SW} r - \left(\frac{3}{2}+\alpha\right) \log \left(a^2 \bar \mu^2 \right)\right)  
\Bigg]
\label{3ptGGgLattT}
\eea
Following the same procedure for extracting the mixing of $S_\mu$ and $T_\mu$ with ${\cal O}_{C7}$, ${\cal O}_{C8}$, we use Eq. \ref{3ptGGgLattS} and Eq.~\ref{3ptGGgLattT}, respectively. We find that there is no mixing with these operators to one loop:
\bea
Z_{SC7}^{LR,\MSbar} &=& Z_{SC8}^{LR,\MSbar} = 0\\
Z_{TC7}^{LR,\MSbar} &=& Z_{TC8}^{LR,\MSbar} = 0
\eea

In contrast, the lattice Green's functions containing gluino-ghost-antighost external fields are identical to the continuum ones in Eqs.~(\ref{3ptGFgCCexpS}) - (\ref{3ptGFgCCexpT}) at one loop order; thus, there is also no mixing either $S_\mu$ or $T_\mu$ with ${\cal O}_{C9}$.

Thus, to one-loop order on the lattice, $S_\mu$ and $T_\mu$ do not mix with class A and class C operators. However, they do mix with the gauge-variant operators of class B, cf. Eqs.~(\ref{ZSB1}) - (\ref{ZTB2}).

\section{Summary and Future Plans}
\label{summary}

In this paper we address the mixing which occurs among the supercurrent operator, $S_\mu$, and the mixing operator, $T_\mu$, beyond tree level with a number of gauge noninvariant operators, using lattice perturbation theory. We employ the Wilson plaquette action for the gluon fields and the Wilson fermion action with the clover improvement for the gluino fields.  

 Extensions of the present work include the application to other actions currently used in numerical simulations, including fermion actions with stout smearing and improved gluon actions \cite{Bergner:2015adz,Ali:2019agk}. In these cases, additional contributions to the renormalization factors are more convergent, and thus their perturbative treatment is conceptually more straightforward; nevertheless, the sheer size of the vertices renders the computation quite cumbersome. 

The results are a first starting point for a nonperturbative calculation of supersymmetric Ward identities and a tuning of the lattice action towards the supersymmetric limit. 
Depending on the method one wishes to employ for computing Green's functions of the supercurrent operator nonperturbatively, a renormalization scheme other than $\MSbar$ may be more appropriate. In particular, one may employ an extension of the X-space scheme, the Gauge Invariant Renormalization Scheme (GIRS), in which conditions need to be imposed on two-point and three-point Green's functions. And in doing so, the new renormalizations and mixing coefficients in GIRS, $Z^{L,GIRS}_{SS}$, $Z^{L,GIRS}_{TT}$ and $Z^{L,GIRS}_{ST}$, $Z^{L,GIRS}_{TS}$ will be related to $Z^{L,\MSbar}_{SS}$, $Z^{L,\MSbar}_{TT}$ and $Z^{L,\MSbar}_{ST}$, $Z^{L,\MSbar}_{TS}$ via a $2 \times 2$ regularization-independent conversion matrix, whose elements are finite functions of the renormalized coupling. In fact, these relevant matrix elements are directly obtainable from continuum calculation. In Ref.~\cite{Bergner:2022}, we aim to present the nonperturbative results in the GIRS scheme along with conversion factors taking us from GIRS to $\MSbar$.

\begin{acknowledgements}
We thank Stefano Piemonte for helpful discussions.
M.C. and H.P. acknowledge financial support from the projects ``Quantum Fields on the Lattice'' and ``Lattice Studies of Strongly Coupled Gauge Theories: Renormalization and Phase Transition'', funded by the Cyprus Research and Innovation Foundation (RIF) under the contract numbers EXCELLENCE/0918/0066 and EXCELLENCE/0421/0025, respectively.
G.S. acknowledges financial support from H2020 project PRACE-6IP (Grant agreement ID: 823767).
G.B. and I.S. acknowledge financial support from the Deutsche Forschungsgemeinschaft (DFG) Grant No.~BE 5942/3-1 and 5942/4-1.

\end{acknowledgements}

\end{document}